# Architecture and Knowledge Representation for Composable Inductive Programming[1]

Edward McDaid FBCS
Chief Technology Officer
Zoea Ltd.

Sarah McDaid PhD
Head of Digital
Zoea Ltd.

## Abstract

We present an update on the current architecture of the Zoea knowledge-based, Composable Inductive Programming system. The Zoea compiler is built using a modern variant of the blackboard architecture. Zoea integrates a large number of knowledge sources that encode different aspects of programming language and software development expertise. We describe the use of synthetic test cases as a ubiquitous form of knowledge and hypothesis representation that supports a variety of reasoning strategies. Some future plans are also outlined.

## 1    Introduction

Computers and software pervade every aspect of modern life. Yet only a small percentage of people are able to program. One of the most significant potential benefits for AI supported software development would be to enable many more people to produce code. This implies that such technology should be accessible to non-technical users and - if it is ever to see widespread adoption – not require significant training.

In modern software development it is common for business analysts, testers, developers and even stakeholders to communicate at least some of their functional requirements using test cases. The prevalence of test-driven development and associated practices make test cases a compelling candidate as the basis for a simple and intuitive software specification language.

It is widely accepted that iterative and incremental approaches to software delivery are beneficial. Few competent developers would expect to produce a large or complex program from a set of requirements in a single iteration. This suggests that any inductive programming process should also be iterative and incremental in nature.

Inductive programming (IP) covers a range of approaches to the automatic generation of code from a specification. Throughout its history inductive programming has been treated predominantly as a technical problem. The above considerations suggest that it also must address problems of representation, process and usability. This allows us to reformulate the core problem of inductive programming in terms of providing systems, notations and processes that allow users to incrementally generate software from a specification. It was in this context that Zoea was created.

---

[1] This paper is an updated and expanded version of McDaid and McDaid [1].

Zoea [1] is a knowledge-based system that incrementally produces software from test cases and without the need for users to learn a complex programming language. A knowledge-based approach provides the ability to build and test components incrementally, inspect and trace reasoning, and to provide explanation. These benefits also help to make construction easier and provide opportunities to further enrich the end user experience. In addition, programming languages and to some extent the process of software development are formal and well understood domains so the domain knowledge is directly accessible to any competent software developer. The domain is also relatively stable in the sense that the core domain knowledge changes relatively infrequently.

Zoea uses a blackboard architecture as a simple and effective means of integrating many diverse knowledge sources. Conventional compilers use a parser and an abstract syntax tree to translate source code into machine instructions. Zoea on the other hand, identifies or hypotheses, data and code elements, based on features in test cases and tries to integrate these in a manner that is similar to a chart parser. There are strong parallels in data terms between charts and blackboards. The benefits of the blackboard architecture, which include a strong domain model and loose architectural coupling, are also highly valued in conventional software development circles.

This paper describes Zoea as it currently exists in terms of its overall approach, architecture, knowledge representation, abstractions, reasoning and operation. To place this in context we will first describe some related work and provide a brief introduction to the Zoea representation languages and associated process.

## 2 Related Work

Inductive programming [2] aims to generate software from a specification such as example data using some form of background knowledge. Automatic programming [3] is a broader field that encompasses many different code generation techniques including compilers and domain specific languages. Program synthesis [4] includes the generation and verification of software through a mathematical description.

A number of inductive programming techniques have been investigated [5]. Currently, approaches based on functional programming [6] and inductive logic programming [7] account for most of the work in this area.

IP systems can also be characterised as enumerative in the sense that they generate and test all possible solutions, or analytical, using algorithms to either avoid some possibilities or directly identify the solution in some way. For example, it is possible to identify some recursive functions directly through analysis of their execution traces [8].

Within each paradigm a wide range of approaches have been investigates but there is also significant overlap between them. For example, generalisation is an important part of program generation to ensure that solutions also work beyond the provided examples.

The term inductive programming is itself a little misleading. Induction as a logical reasoning technique is sometimes used but other forms of reasoning including deduc-tion and abduction are also employed. Techniques used include various forms of search, inference, Bayesian networks, genetic algorithms and grammar-based approaches. Machine learning has also been applied to programming by example, notably to learn associations between features of the input data and generated programs in the training set [9].

The combinatorial explosion is a significant problem for all IP systems. Generate and test is directly affected but analytic approaches are also impacted indirectly. As a result even the most advanced IP systems are currently able to produce relatively small programs [10].

Often the target language is the same as the IP system implementation language although there is nothing to preclude the design of a new target language that is optimised in some way to support IP. It has also been suggested that IP could benefit from the use of extensive function libraries [11].

Machine learning has been applied to create systems that provide intelligent auto-completion for source code being edited in an IDE [12]. Typically these are trained on a large corpus of existing code to create a model that can be used for prediction. Similar approaches have been used to also produce systems that allow people to describe programs using natural language [13].

Modern programming languages provide developers with a wide range of facilities. However, not all of these are used with equal frequency. Analysis of large code repositories has provided a better understanding of how developers actually use programming languages [14].

The blackboard architecture has a long history in AI [15] and many variants have been developed [16]. As an architectural pattern the blackboard model remains the simplest way to integrate multiple loosely coupled components around a central data model. Similar approaches are widely used beyond AI (although rarely credited) as the core of many high volume distributed applications. There are also a number of closely related technologies including event stream processing, tuple spaces and chart parsers.

It has been suggested that both deep learning and symbolic approaches to AI could benefit from being more closely integrated [17]. The blackboard architecture provides an ideal framework for such heterogeneity. Indeed, it has been proposed that the human brain may use a similar model to integrate the processing of different cortical regions in the thalamus [18].

## 3    Zoea Approach

One of the key problems faced by inductive programming approaches is the combinatorial explosion. For example, the generate-and-test approach is relatively simple to implement, but quickly results in enormous numbers of hypotheses as program size increases. There are a number of ways in which this problem can be mitigated but not eliminated. As a result brute force search alone will always struggle to produce anything beyond relatively small programs.

Best first and related strategies can be useful in some scenarios, for example where a target program involves string manipulation. This requires a suitable distance function. Best first search is used in some Zoea knowledge sources and q-gram distance [19] was found to be the most effective distance measure. Information entropy and string length metrics were also evaluated but provided no significant benefits over edit distance measures.

One important characteristic of how Zoea generates code it is that the focus - at least initially - is on producing values rather than generating code. In Zoea code generation is a two-stage process. The first stage involves forward chaining from an input or an intermediate value to produce new intermediate values using suitable target language instructions. The second stage only occurs once a target value has been generated. In this phase the corresponding code is generated by chaining backwards from the target value to the inputs. This allows the data flow graph for a code fragment to be created by identifying any single data flow path from the input to the output. The other data flow paths are then grown from intermediate values like aerial roots that occur in some plants. Deltas are used to identify any additional values that are not already derived and an attempt is made to generate code for these also.

Zoea uses a knowledge-based approach that encodes expertise about various programming language elements and also how they can be combined in different ways. Pattern recognition is also used to extract as much information about the required program as possible from the test cases.

When a human developer is presented with a set of test cases they can often identify characteristics of the corresponding program. This is because the test case input and output values can provide important clues about the code that they describe. For a start the data types can indicate directly in some cases that specific type conversions may be required. For example given a single character as an input and a small integer as the output many programmers would suspect that the number was an ASCII code or similar encoding. The presence of the same value in both the input and output of a test case may be a coincidence but it is also evidence to suggest that that value may simply be mapped directly. Similarly, if all of the output elements are already present in the input then this suggests that the required code may just select information or move it around rather than creating any new values.

The actual values of inputs and outputs can also be useful. For example, recognising a problem as string manipulation as opposed to being mathematical in nature makes identifying the required program more tractable. People are also able to interpret data in multiple ways such as a string that contains a JSON data structure, which describes a collection of objects. Simply by inspection people are often able to determine if an input and output pair of lists has been sorted, reversed, filtered or selected in any combination. Zoea is designed to take advantage of the fact that a user is actively trying to communicate their requirements through the test cases that they create.

The granularity of knowledge representation within a knowledge-based system is an important consideration. Zoea takes a coarse grain approach in which knowledge that applies to language elements is as generic as possible. For example, consider a knowledge source that is concerned with mapping of data elements between different types of composite data structures. It would be possible to define separate rules for mapping between each combination of lists, arrays, collections and hash-maps - potentially with different numbers of dimensions. In Zoea this is handled by a single knowledge source using a standardised path-based addressing scheme for all data.

Each time a knowledge source detects a pattern in a test case input it is effectively creating a hypothesis. For example if the Date knowledge source identifies a string in a test case that looks like a date then there are two possibilities. The string could be a date or it could just be a string after all. Zoea captures and manages such hypotheses by creating synthetic test cases. A synthetic test case is no different to a real test case except in its provenance. In our date example the Date knowledge source creates a synthetic test case that is the same as the original test case except that the string that looks like a date is converted to an actual date. The knowledge source also creates an associated code fragment that converts the data in the original test case into a date. Synthetic test cases are then processed in exactly the same way as real test cases which may include the creation of additional synthetic test cases and so on.

Synthetic test cases allow Zoea to explore different interpretations of the test cases it has been provided. For example consider a test case whose input and output are equal length arrays of numbers. One possibility is that there is a one-to-one correspondence between the input and output array elements. This possibility is explored through the creation of a set of synthetic test cases consisting of the nth input and output array elements respectively. At the same time Zoea considers other possible interpretations of the original test case. If one or more code solutions for the synthetic test case are found then these are integrated into the code that iterates the input array.

The combination of pattern matching against test case data and relatively coarse grained programming knowledge (where application of that pattern is at least plausible in the current context) help to guide Zoea towards producing partial and complete solutions that reflect the users intention as articulated through the test cases.

## 4    Composable Inductive Programming

Composable inductive programming (CIP) [20] is a novel software development paradigm. It is a simple, iterative process in which all software is produced through inductive programming. Users provide a specification that consists of a set of test cases. Larger programs are assembled (or composed) by combining smaller programs in various ways. This allows CIP to be used to create software of any size.

Test cases usually include input and output values. Zoea also supports any number of intermediate (or derived) values. Derived values allow the user to describe complex computations within a test case. Internally Zoea treats derived values as representing multiple input-output test cases. For example, a single test case that takes the form input-derive-output is equivalent to two separate test cases corresponding to input-derive and derive-output respectively.

Zoea also allows any number of existing programs that have already been compiled to be combined to form larger program. This form of composition is called 'use' and it is similar in concept to 'include' in conventional programming languages except that there is no indication of where or how the used program should be integrated into the solution code. Zoea treats use as a suggestion that it should try to include the named program in the solution for the current program, if it can.

In terms of the CIP operating model a user begins by producing a set of test cases using one of the two available Zoea notations (described below) corresponding to the program they want to generate. Complex problems should be broken down into smaller programs that can be worked on separately.

The test cases are submitted to the Zoea compiler for compilation. If compilation is unsuccessful in the sense that Zoea is unable to produce a code solution within an acceptable length of time then the user must modify or extend the test cases. If compilation is successful then the user should test the generated program to ensure correct behaviour. The process can iterate any number of times using composition if necessary to combine smaller programs as required.

## 5 Zoea Specification Language

Zoea Specification Language (ZSL) [20] is a simple language used to describe programs as a set of test cases. A ZSL program contains one or more terms where each term has the form <Tag> : <Value>. Terms must be separated by white space but layout is not significant.

ZSL defines the following tags: program, use, data, case, step, input, derive, output. The program tag is always required. The optional use tag can identify one or more existing programs to include. An optional data tag allows static reference data to be included in a program.

One or more cases are identified by the case tag, which is optional for a single case. Each case can have any number of optionally identified steps where a step contains a single input, derive or output tag. Data, input, derive and output values are basically JSON data although quotes are optional for strings that contain no special characters.

Derived values in ZSL are effectively an abstraction over core Zoea test cases which consist only of an input and an output value. ZSL supports only single values, which may be composite. If multiple input or output values are required then these must be represented by convention using a composite data structure.

## 6 Zoea Visual

Zoea Visual [21] is a visual programming language for describing Zoea test cases that is built on top of ZSL (see Fig. 1). In Zoea Visual all data is represented visually as input fields, tables, etc. Elements are placed in named columns that correspond to ZSL tag names. Zoea Visual supports any number of input and output values.

Zoea Visual introduces the concept of dependencies, which are data flow links between pairs of elements in different columns. These are used to indicate that one or more source elements are involved in some way in the production of a target element. This allows the user to construct a data flow diagram for each test case that provides more detail about the user intention to the compiler.

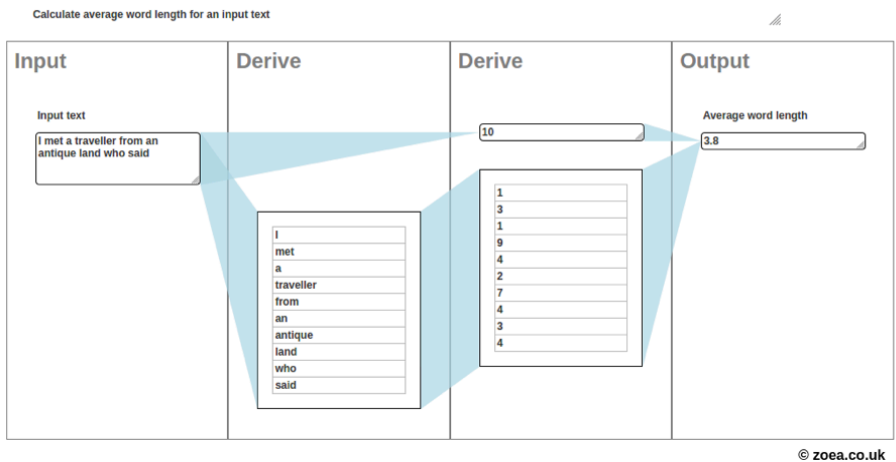

**Fig. 1.** Example Zoea Visual Program

Zoea Visual also introduces a further form of composition called subsidiary test cases. A subsidiary test case allows the code for a given element to be elaborated as one or more completely separate Zoea Visual programs. This facility allows complex logic and multiple code paths to be expressed using a linear rather than exponential number of test cases.

## 7    Zoea Architecture

Zoea uses a version of the blackboard architecture in which many knowledge sources update and react to changes in a central data store (see Fig. 2). There are a number of differences between the Zoea blackboard and the classical model:

- Knowledge sources are distributed across a number of physical computers;
- Knowledge sources work on local copies of relevant subsets of the blackboard that are synchronised in both directions through deltas;
- The central copy of the blackboard is mainly used for data integration and redistribution;
- Global scheduling relates exclusively to the allocation of time and resource budgets - time is elapsed time in seconds managed via time-to-live counters and resources are processing capacity expressed as CPU/seconds;
- Knowledge sources act concurrently within their allocated time and resource budget;
- Control knowledge that relates to specific knowledge sources is part of the knowledge source rather than an external scheduler.

The knowledge sources are built on top of a distributed computing framework that allows arbitrary compute jobs to be unicast or multicast to a pool of workers located on any number of physical nodes. A central controller monitors the activity of all workers, and distributes jobs including knowledge source activations and blackboard updates as required. A knowledge source activation is simply the allocation of a knowledge source to a test case.

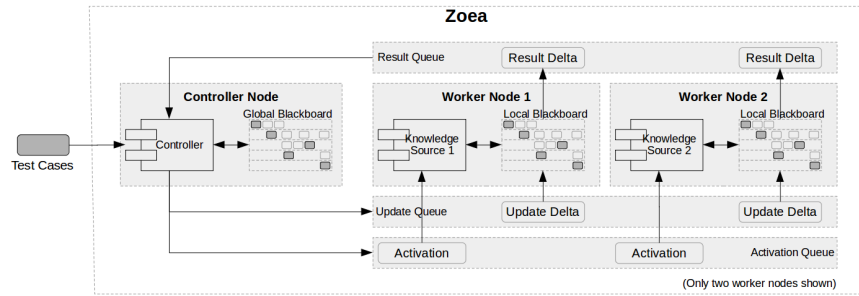

**Fig. 2.** Zoea blackboard architecture

At the test case level the overall behaviour is breadth first search driven by a queue of test case/knowledge source activations. Knowledge sources often give rise to additional synthetic test cases, which are linked to the originating test case to form a hierarchy. All abstraction levels are partitioned by test case identifier and test case number. This allows Zoea to compile multiple programs concurrently and also enables the blackboard to be invoked recursively with respect to synthetic test cases.

It is possible for more than one synthetic test case to be created with exactly the same input and output data. Such situations are detected by the controller, which only allows the first instance of any such set to be executed. In all subsequent occurrences knowledge source execution is bypassed and instead any results from the first instance are mapped to the relevant abstraction levels of the current synthetic test case.

All Zoea Visual constructs are translated to ZSL for compilation. Separate ZSL programs are created for each set of dependencies with the same target and for each derive/output element with no dependencies - both for the root diagram and also for every subsidiary test case.

Zoea also includes a number of tools to support knowledge source development, test case minimisation, debugging and tuning. Knowledge sources can be individually enabled and disabled. Activity and execution timings are logged during execution. Aggregate counts of all blackboard entities are also recorded at key lifecycle events. An extensive regression test suite of ZSL code with known and manually verified results is run against every revision of the Zoea codebase. Zoea has a limited but useful ability to tune itself automatically so as to minimise execution time for the regression suite and maximise worker utilisation.

In Zoea the controller, all knowledge sources and all workers are single threaded. This obviates any need for thread synchronisation, copying data across threads and reduces overall indeterminism.

# 8 Knowledge Representation

In Zoea test cases are used as the focal point for all problem solving. At the highest level the test cases that are provided by the user express the problem of determining the code that will transform each test case input into the corresponding output. The knowledge sources address this problem by identifying patterns in the test case data. If a pattern is found then the knowledge source will generate some code and it may also produce a different synthetic test case, which is then treated as a completely new problem. Should the synthetic test case eventually yield a solution then the knowledge source modifies and integrates it with the code fragment it has already created.

The Zoea blackboard is organised as a number of abstraction levels which represent a model of the domain that is used to support problem solving. Each abstraction level corresponds to a set of entries of a given type. Entries can have any number of associated typed attributes, including typed links to other entries on any abstraction level. All entries have a unique identifier.

In Zoea the abstraction levels reflect the problem of translating a set of test cases into a program that works with all of the test cases. During compilation each test case is treated separately so while some of the abstraction levels such as the solution will apply to all of the test cases for a given program, others apply to a specific test case only.

Between the test cases and solutions the abstraction levels reflect the general stages of the problem solving process:

- Identifying features in test cases;
- Applying test cases to knowledge sources;
- Creating new values from existing values;
- Finding code fragments for a value;
- Finding partial solutions for a single case;
- Finding complete solutions for a single case;
- Finding a set of case solutions that work for all cases;
- Generating the raw code for a solution set;
- Generalise the raw solution set code.

The entries on the Zoea blackboard do not form a simple linear hierarchy. Rather, the creation of synthetic test cases and the knowledge source activations form a large tree in which the code elements of any solution are both sparsely scattered and also mutated as a result of synthetic test case transformations. (See Fig. 3) It is the responsibility of each knowledge source to collect the solutions for the synthetic test cases it has created, and to transform and integrate the associated code. Partial and complete solutions include metadata that identifies their origin in terms of synthetic test case and knowledge source.

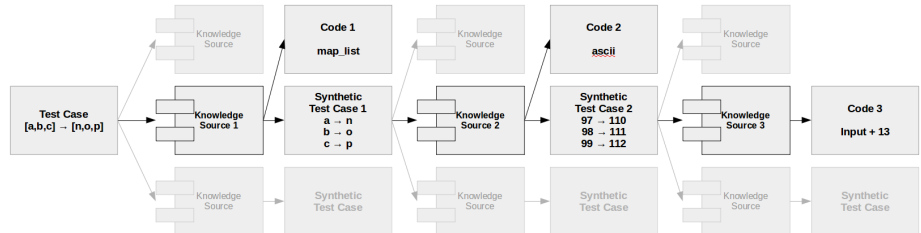

**Fig. 3.** Knowledge sources and synthetic test cases

The blackboard entries for individual knowledge sources are also isolated from one another by an additional layer of synthetic test cases. This is accomplished by the controller, which creates multiple identical copies of each synthetic test case for each knowledge source. The overhead in creating and managing the additional test cases is very small but the benefit is the elimination of any potential for cross talk between knowledge sources.

## 9 Knowledge Sources

From a black box perspective a Zoea knowledge source takes a synthetic test case as input and produces code and/or another synthetic test case as output. In a sense Zoea operates in a manner that resembles a rewriting system that works on test cases.

All knowledge sources are invoked with respect to a particular synthetic test case identifier. Given this they are able to access any other information they may need directly from the blackboard. In each case the knowledge source is responsible for determining whether it is applicable.

All knowledge sources are composed of a standard set of code elements. In addition to this they are always invoked by a wrapper that provides a standard set of execution related services.

A Zoea knowledge source includes some or all of the following elements:

- Knowledge source specific unit tests (mandatory);
- Resource manager (mandatory);
- Pre-condition (mandatory);
- Pattern matcher (mandatory);
- Child synthetic test case producer (optional);
- Child synthetic test case result extractor (optional);
- Child synthetic test case code mapper (optional);
- Solution code generator (mandatory);
- Solution verifier (mandatory).

The knowledge source wrapper maintains a global call stack of current knowledge source activations. This allows knowledge sources to determine whether they have been called recursively. The wrapper is also responsible for knowledge source level logging, duration measurement and data cleanup.

## 10    Reasoning

Synthetic test cases capture the hypotheses that are created by the knowledge sources. In this sense their use approximates to a hypothesis testing problem-solving strategy. Each set of synthetic test cases is different in some way from the test cases from which they were derived. Depending on the knowledge source the input and output elements of the source test cases are transformed through some combination of mapping, mutation and decomposition. This corresponds to problems being decomposed into simpler problems and/or transformed into different problems.

Each test case that relates to a particular program corresponds to a single code path in the solution. When Zoea identifies a solution that works for one case it also tries the same code with every other test case. As a result the various knowledge sources may produce many different case solutions and each of these will have an associated list of test cases to which they are applicable.

If one or more case solutions exist that are applicable to all test cases then Zoea will select the simplest of these. Solution complexity is based on a variant of Halstead length [22] that is weighted for the presence of literal values of different data types.

Many target programs will include multiple code paths and as a result all case solutions will be applicable only to a subset of test cases. Rather than representing the complete solution the case solutions each may correspond to part of a conditional statement. In order to produce the solution code it is necessary first to identify a set of case solutions that covers all of the test cases. This is a set-cover problem. Second we must determine the conditional logic required to differentiate between the test cases expressed in terms of input or derived values. Typically very many solutions can be generated for a set of case solutions so preference is given to those that include the smallest number of case solutions. Again the simplest solution code from all of those produced is selected.

Identifying a code solution that accounts for a set of test cases is an example of abductive reasoning. The code is the simplest explanation for a set of observations in the form of the test cases. The process of selecting a subset of case solutions that collectively address all of the test cases is an instance of set-cover abduction. The same process and mechanism is widely used in Zoea, for example in the creation of logical expressions.

## 11    Discussion and Future Work

ZSL and Zoea Visual are both novel, highly declarative programming languages. Composable inductive programming also represents a new software development paradigm.

Most of the Zoea code examples that have been published are relatively simple. Internally both ZSL and Zoea Visual have been used to manually create programs that are the equivalent to hundreds of lines of code in conventional languages. In addition, automated testing with randomised programs has produced hundreds of thousands of programs - some of which are equivalent to over a thousand lines of conventional code.

While the compiler is functional work continues to improve performance and to provide more end user tooling. In particular we want to make hypotheses visible to the user, allowing them to be optionally accepted or rejected. We also plan to provide some accessible representation of the generated software rather than source code. In addition we have a long backlog of features for ZSL and Zoea Visual which includes support for file, database and network operations.

Compilation time depends on the complexity of the target program. For the regression suite this varies between under a second up to around two minutes with a generally reciprocal distribution. Increasing the number of workers reduces compilation times.

As a knowledge-based system development of Zoea has been relatively straightforward. The domain of programming languages and software development is nicely formal and deterministic with little of the uncertainty, brittleness and rule interaction that can be found in other problem areas.

While the knowledge sources vary considerably in size and complexity their development is otherwise relatively straightforward. Originally all knowledge sources worked in the same synthetic test case context. This caused a lot of complexity even though the metadata was there to disentangle the entries. The introduction of knowledge source isolation made development more straightforward.

The synthetic test case transformations carried out by the knowledge sources could be defined in more abstract terms. This might be useful for reasoning about interaction between knowledge sources for example to eliminate possibly redundant combinations or to anticipate possible activations in advance. It may also be useful as some form of completeness check.

Early versions of the Zoea blackboard devoted an excessive amount of effort to data synchronisation. This became apparent with the introduction of performance instrumentation. In the current version knowledge sources get access only to the information they actually require.

Currently Zoea is run on a dedicated cluster of physical servers. While this is a cost effective model for development and testing it limits access for other purposes. At some stage we would like to make Zoea more publicly available and to this end it is currently being ported to cloud-based infrastructure.

## 12    Conclusions

Zoea is the first inductive programming system that was designed to allow the creation of code of arbitrary size. We have outlined the knowledge-based approach and blackboard architecture used to build it. Zoea uses knowledge-based pattern recognition to identify opportunities to apply knowledge and it uses synthetic test cases to manage hypotheses. In giving rise to very simple, novel programming languages and processes, Zoea demonstrates that coding can be made much easier than it currently is.

**Acknowledgements**

This work was supported entirely by Zoea Ltd (https://www.zoea.co.uk). Zoea is a trademark of Zoea Ltd. All other trademarks are the property of their respective owners. Copyright © Zoea Ltd. 2022. All rights reserved.